\documentclass{ws-ijmpe}
\usepackage[super,compress]{cite}
\usepackage{epsfig}
\usepackage{multirow}
\begin{document}



\title{Multiplicity Moments using Tsallis Statistics in High-Energy Hadron-Nucleus Interactions}
\author{S. Sharma* $^{1,2}$, G. Chaudhary$^1$, K. Sandeep$^1$, A. Singla$^1$ and M. Kaur$^1$ }
\address{1. Department of Physics, Panjab University, Chandigarh\\
2. CGC-College of Engineering, Landran, Mohali\\
India\\
sandeep.sharma.hep@gmail.com\footnote{corresponding author}}
\maketitle
\begin{abstract}
The study of higher-order moments of a distribution and its cumulants constitute a sensitive tool to investigate the correlations between the particle produced in high energy interactions. In our previous work we have used the Tsallis $q$ statistics, NBD, Gamma and shifted Gamma distributions to describe the multiplicity distributions in $\pi ^-$ -nucleus and $p$ -nucleus fixed target interactions at various energies ranging from P$_{Lab}$  = 27 GeV to 800 GeV. In the present study we have extended  our analysis by calculating the moments using the Tsallis model at these fixed target experiment data. By using the Tsallis model we have also calculated the average charged multiplicity and its dependence on energy. It is found that the average charged multiplicity and moments predicted by the Tsallis statistics are in much agreement with the experimental values and indicates the success of the Tsallis model on data from visual detectors. The study of moments also illustrates that KNO scaling hypothesis holds good at these energies.

\end{abstract}  
\ccode{PACS numbers: 5.90.+m, 13.66.Bc}

\section{Introduction}
Collisions of particles at relativistic high energies lead to the production of various new elementary particles. These particles are produced due to the gluon-gluon, quark-quark and quark-gluon interactions\cite{int1} between the constituent quarks and gluons of the colliding particles. These collisions can be hadronic\cite{hh2} , leptonic\cite{ee3} or hadron-nucleus\cite{ha4} in nature. Among these interactions study of hadron-nucleus ($hA$) interactions plays a significant role in understanding the mechanism of hadron production and their properties. In these interactions nuclear fragmentation products reflect in their characteristics and the mechanism of  production of new particles\cite{fragmentation5}. It is quite promising to investigate the correlation between the various types of particles produced in the final state of $hA$ collision. The study of high energy hadron-nucleus interactions  become very important to understand the particle-particle interactions and the phenomenon of particle production\cite{e0} in heavy ions interactions in nuclear targets. Heavy ion interactions play a key role in the understanding of physics of formation of quark-gluon plasma (QGP).\cite{qgp6} Nucleus-nucleus interactions can be explained as a superposition of hadron-nucleus interactions. Particles produced in the $hA$ interactions are studied using various phenomenological models. 
 The study of charged  multiplicity in the final state of high energy interactions can unveil information about the series of events that occur at the early stage of interaction. \cite{md7} Analysis of the charged particle multiplicity gives an understanding about the dynamics of formation of hadrons as a combination of quarks (anti-quarks) and gluons, collectively known as partons. Charged particle production in final state exhibits the footprints  of this evolution of hadrons from the partons, embedded in the form of correlations among the particles. The absence of any kind of correlation among the particles gives the Poissonian form of multiplicity distribution. \cite{poisson8} On other hand, if the production of one particle intensifies the production probability of other particles, then the distribution deviates from Poissonian form. The multiplicity moments become meaningful when one needs to study the properties of multiplicity distributions. The correlation between the produced particles can be studied precisely using higher order moments and their cumulants. \cite{cummulants9} 
 
 Various  models and distributions\cite{models10} have been used to make predictions for charged particle multiplicity which uses the concepts of probability and statistical mechanics. The statistical mechanics has the basic assumption that the quantities like internal energy, volume and Gibbs-entropy  are extensive in nature. However, this is normally  acceptable when we study and analyse the short range interactions. But when we analyse the long range interactions like quark-quark, gluon-quark and gluon-gluon interactions this assumption no longer stands valid. In such interactions the standard statistical mechanics which is extensive, becomes non-extensive in nature. Therefore, it becomes very crucial to take care of the non-extensive nature during the study of important observables like multiplicity  and transverse momentum.\cite{pt11}
 
 Constantino Tsallis\cite{Tsallis12} introduced a possible and desirable solution for this problem. He brought forward the concept of replacing the regular Gibbs entropy with a new Tsallis entropy which is non-extensive in nature. This Tsallis entropy is indexed by a parameter $q$, a real-valued parameter, which measures the extent of deviation from extensivity. Most of the results from statistical mechanics can be transformed into  this new concept. In study of production of quark-gluon plasma  in heavy-ion collisions at higher energies, thermostatistics  is notably significant. In very high energy collisions, statistical equilibrium  is supposed to be achieved which leads to the non exponential form of transverse energy distribution of the hadrons produced during the collisions.  This non exponential behaviour has been observed not only in heavy ion collisions but also in the leptonic collisions, $e^+ e^- \rightarrow hadrons$  as well as in hadronic collisions, $pp \: (\bar{p} p) \rightarrow hadrons$ and can be described very well by adopting non-extensive equilibrium as ascribed to the Tsallis non-extensive thermodynamics.  In above mentioned scenarios, the Tsallis statistics is the best possible solution and  technique available to describe the multiplicity distributions and transverse momentum distributions for a broad range of energies. The Tsallis non-extensive phenomenon, thus plays a key role in the high energy collisions and this non-extensive part of entropy is evaluated by parameter $q$, the entropic index in the Tsallis function. \cite{kodoma13} In this case the total Tsallis entropy of two sub-systems a and b is given by;
 \begin{equation}
 S_q(a,b)= S_a + S_b + (1-q)S_{a}S_{b}\, \label{one}
 \end{equation}

Over the past few years, the Tsallis model was widely used in analysing the MDs in leptonic, hadronic and heavy ion collisions and found to be most successful in describing the experimental data. In the last few years we have also used the Tsallis model to analyse the data from $e^{+}e^{-}$, $\overline{p}p$ and $pp$ collisions for a broad range of energies  from lower energies (GeV scale) up to the highest available  energies at the LHC.\cite{s1,s2,s3,s4}  In one of our previous analyses\cite{mypaper17} we have used the four approaches namely, negative binomial distribution \cite{nbd14} , gamma distribution \cite{gamma15} , shifted gamma distribution \cite{shifted gamma16} and the Tsallis q-statistics \cite{kodoma13} to analyse hadron-nucleus data from fixed target experiment. In that analysis comparison between the distributions were done on the basis of minimal $\chi ^2$ but none of these distribution could emerge as clear cut choice. ~The lack of clear cut choice of distribution to describe the experimental data  and success of the Tsallis model for a wide range of energies motivate us to further extend our analysis on fixed target experiment data.  In this current study we have used the Tsallis model to analyse the correlations between the particles produced in 
$\pi ^-$-nucleus and $p$-nucleus fixed target interactions at various energies P$_{Lab}$ ranging from 27 GeV to 800 GeV  to affirm our conclusions. 

In Section 2, concise information on moments, Tsallis distrbution, formulation to calculate higher order moments and calculation of uncertainity on moments are given.~ Section 3 gives the details of data used in current work. Section~4 presents the analyses of experimental data and the results obtained.~Discussion and conclusion are presented in Section~5.
\section{Moments}
Multiplicity distribution carries important information about the the correlations of the particle produced in the particle collision processes. The shapes of the MDs vary highly for different processes at various energies. The distribution cannot be described only by the possibilities of N particles generated in the process but also by the generating function of the probability distribution function.\cite{generating18} The multiplicity distribution obeys conventional Poisson distribution if there is no correlation between the particles produced as in the case of interactions  at lower energies. In this scenario dispersion  is related with the average multiplicity $<N>$ as $D = \sqrt{<N>}$. Deviations from a Poisson distribution signifies the correlations amongst the produced particles. Using the presumption that the average multiplicity could be used to formulate the energy dependence of multiplicity distribution at higher energies, the shape of any multiplicity distribution can be outlined. In 1972 Koba, Nielsen and Olesen proposed the theory of universal scaling \cite{kno19} , called as KNO scaling,  for multiplicity distributions at high energies to address the matter of energy dependence of multiplicity. The energy dependence at high energy collisons approximated  by relation $D$ $\propto$  $<N> $ implied the compliance of KNO scaling. To study the validity of scaling hypothesis higher order moments and its cumulants portray very important role. The normalized moments $C_\alpha$ of order $\alpha$ play a key role in determing the violation or holding of KNO scaling. If scaling hypothesis holds good then the normalized  moments should be independent of the energy. Not only the validity of KNO scaling but the correlation between the produced particles  can also be studied and understood correctly by using factorial moments, $F_\alpha$ of order $\alpha$. In case of Poisson distribution the factorial moments\cite{m20} comes out to be 1 for all values of $\alpha$  because of independent particle production . These moments are defined as; 

\vspace{-0.5cm}
\begin{center}
 \begin{equation}
C_\alpha = \frac{<N^\alpha>}{<N>^\alpha}  
   \end{equation}
 \end{center}
 \vspace{-0.5cm}
 
\begin{center}
 \begin{equation}
F_\alpha = \frac{<(N(N-1)....(N-\alpha +1))>}{<N>^\alpha}
\end{equation}
 \end{center}
 \vspace{-0.2cm}
 
 The factorial moments and their cumulants, $K_\alpha$, are approximately decisive in describing the tail part of distribution\cite{m21} where events with multitude of particles give a meaningful contribution. The factorial moments and cumulants are related to each other by the relation;
 \vspace{-0.7cm}
\begin{center}
 \begin{equation}
F_\alpha = \sum _{i=0} ^{\alpha-1} C _{\alpha-1} ^i K_{\alpha-i} F_i
\end{equation}
 \end{center}

  Factorial moments emphasize the presence of correlation amongst the particles and cumulants of order $\alpha$ exhibit complete $\alpha$-particle correlation which can not be brought down to the lower order correlation. In precise manner it means, if all $\alpha$ particles are related to each other in $\alpha^{th}$ order of cumulants, then it can not be splitted in to disconnected groups i.e. $\alpha$ particle cluster can not be divided in to further smaller clusters\cite{m, m22}. The study of moments and their dependence on energy helps in formulating, revisting, accepting and rejecting various statistical  and Monte-Carlo models  used in describing the partcle production mechanism. 

\subsection{Tsallis Distribution}

Tsallis modified the usual Boltzman-Gibbs formulation \cite{gibbs23} to introduce  the concept of non extensive behaviour of entropy. This non-extensive entropy is given by;
\begin{equation}
S = \frac{1-\sum_{a}P_{a}^q}{q-1}\, 
\end{equation}
where the probability associated with microstate $a$ is given by $P_a$ and sum of the probabilities over all microstates is normalized with $\sum_{a}P_{a}=1$. The entropic index $q$ with value $q>1$ and $1-q$ evaluates the variation of entropy from its extensive behaviour.

In Tsallis $q$-statistics, the probability distribution function ($pdf$) is illustrated by the partition function $Z$ and defined as,
 \vspace{-0.2cm}
\begin{equation}
P_N = \frac{Z^{N}_q}{Z}
\end{equation}
where $Z$ represents  the total partition function and $Z^{N}_q$ represents partition function at a particular multiplicity, of the grand canonical ensemble of gas consisting of N particles. For N particles, partition function is expressed as;
 \vspace{0cm}
\begin{equation}
Z(\beta,\mu,V) = \sum(\frac{1}{N!})(nV-nv_{0}N)^{N}
\end{equation}
$n$ represents the gas density, V is the volume of the system and $v_{0}$ is the excluded volume.
The generating function of the distribution plays a significant role in determining the physical information of the multiplicity distribution. The generating function for a given multiplicity distribution which is related with the probability can be written as; 
\vspace{-0.5cm}
\begin{center}	
\begin{equation}
G(t) \: = \: \sum_{N=0}^{\infty} \: P_N t^N 
\end{equation}
 \end{center}
The generating function of the Tsallis distribution can be obtained by using the expression of  probability distribution function and is given by;
 \vspace{-0.5cm}
  \begin{center}
  	
\begin{multline}
G(t) \approx \exp(t-1) V n [1 + (q-1) \lambda (V n \lambda - 1) -2 v_0 n ]\\
+ (t-1)^2 (V n)^2 [(q-1) \frac{\lambda ^2}{2} - \frac{v_0}{V} ]
\end{multline}
 \end{center}

The generating function of Tsallis probability  has the same form as that of Negative Binomial distribution ($G_{NBD} = [1 - \frac{<N>}{k} (t-1) ]^{-k}$ = $\exp [<N> ( t-1)]$) with average of number of particles $\bar{N}$ for Tsallis probability as;
 \begin{equation}
 \bar{N}=Vn[1 + (q-1)\lambda(V n\lambda-1)- 2v_0n], 
 \end{equation} 
 where $\lambda$ is related to the temprerature through parameter $\lambda$ as; 
  \begin{center}
  \vspace{-0.4cm}
 \begin{equation}
\lambda (\beta, \mu)  = - \frac{\beta}{n} \frac{\partial n}{\partial \beta}
\end{equation}
 \end{center}

More details and the method to calculate the partition function for $N$ particles, in the Grand Canonical Ensemble, Tsallis $q$-index and Tsallis $N$-particle probability distribution can be found in reference [\refcite{kodoma13}].

\subsection{ Higher order Moments}

Using the average number of particles produced, we can calculate the normalized moments of order $\alpha$ for the Tsallis disribution. These moments are given by $C_\alpha = \frac{<N^\alpha>}{<N>^\alpha}$, with average multiplicity $<N>$ ($\bar{N}$ = $<N>$). The factorial moments are defined as;
 \vspace{-0.3cm}
\begin{center}
 \begin{equation}
F_\alpha = \frac{<(N(N-1)....(N-\alpha+1))>}{<N>^\alpha}
\end{equation}
 \end{center}
 
 \begin{center}
 \begin{equation}
F_\alpha = \Big(\frac{1}{<N>^\alpha}\Big) \frac{d^{\alpha} G(t)}{dt^\alpha}
\end{equation}
 \end{center}

 with $G(t)$ is the generating function of the Tsallis distribution given by equation (9). The factorial moments and normalised moments are related to each other and can be expressed in the form of $C_\alpha$. The first five factorial moments are;
  %
\begin{multline}
\hspace{3.7cm}F_2 = C_2 - \frac{C_1}{<N>} \\ 
F_3 = C_3 - 3\frac{C_2}{<N>} + 2\frac{C_1}{<N>^2} \\
\hspace{2cm} F_4 = C_4 - 6 \frac{C_3}{<N>} + 11\frac{C_2}{<N>^2} - 6\frac{C_1}{<N>^3}  \\
\hspace{3.7cm}  F_5 = C_5 - 10 \frac{C_4}{<N> } + 35\frac{C_3}{<N>^2} -  50\frac{C_2}{<N>^3} +  24\frac{C_1}{<N>^4} \\
  \end{multline}
\subsection{ Uncertainity on Moments}
The method of partial derivatives is used to calculate the errors on the moments for
 a specified normalized probability distribution $P_{N}$, possessing an uncertainity $e_{N}$. This method is reliable with the presumption that errors on the individual bins possess no  correlation. \cite{error24} Using this method, 
 
 \begin{equation} 
\frac{\partial C_\alpha}{\partial P_N} = \frac{N^{\alpha}\langle N \rangle - \langle N^{\alpha} \rangle \alpha N}{\langle N \rangle ^{\alpha+1}}
\end{equation}
\begin{equation}
\frac{\partial F_\alpha}{\partial P_N} = \frac{N(N-1)....(N-\alpha +1)\langle N \rangle - \langle N(N-1)......(N-\alpha+1)\rangle \alpha N}{\langle N \rangle ^{\alpha+1}} 
\end{equation}
The total is then
\begin{equation}
E_{\alpha}^2 = \sum_{N}\Big(\frac {\partial X_{\alpha}}{\partial P_N}e_N \Big)^2 
\end{equation} 
where $X_{\alpha}$ is $C_{\alpha}$ or $F_{\alpha}$

\normalsize

\section{Data used}
In present study the experimental data from pion-emulsion and proton-emulsion interactions from different fixed target experiments from the energy range 27 to 800 GeV have been analysed \cite{mypaper17}. It is interesting to revisit these old  data for the Tsallis model which successfully describe the present day data at higher energies. The data under study on pion-nucleus interactions mainly come  from the fixed target experiments using nuclear emulsions as the detector \cite{fixexperiment25}. The passage of a charged particle through nuclear emulsion leaves behind a trail of ionization produced in AgBr crystals which are reduced to specks of silver. This trail of specks, known as track, is scrutinized under high power precision telescopes. In the present study $\pi^-$-Em data at $P_{Lab}$ = 50 GeV, 200 GeV, 340 GeV and 525 GeV  have been analysed. Along with the $\pi^-$-Em data, we have also analysed the data of proton-Emulsion, $p$-Em  interactions at $P_{Lab}$ = 27 GeV, 67 GeV, 200 GeV, 300 GeV, 400 GeV and 800 GeV. More details about the data can be found in the reference [\refcite{mypaper17}].

\section{Analysis and Results}
In this section results of the average charged miltiplicities and higher order moments calculated using the Tsallis q-statistics and their comparison with the experimental data are discussed.
\subsection{Average multiplicities}
The probability distribution functions calculated from the Tsallis models have been used for the  $\pi^-$-Em and $p$-Em experimental data. For the Tsallis model the non-extensive parameter $q$ in every case is found to be greater than 1 and increases linearly with increase in energy, $P_{Lab}$ as discussed in our previous analysis. One important feature of multiparticle states in high energy interactions is evolution of mean charged multiplicity with energy\cite{meann26, meann27}. In the present analyses it is observed that dependence of average charged multiplicity can be described well using thes power law. The variation of the mean multiplicity as a function of energy, $P_{Lab}$  follows the power law $<N>$ = $a \; (P_{Lab})^b$ . The $q$ values and average multiplicity values calculated from  the Tsallis model are given in Table I. These values  are compared with the experimental values for $\pi^-$-Em and $p$-Em interactions and found to be in good agreement as shown in figures 1 $\&$ 2. In the present analysis, the power law which describes the dependence of average multiplicity,$<N>$ on energy $P_{Lab}$ takes the form as follows, 
\vspace{1cm}

For $\pi^-$-Em;
\begin{center}
\begin{equation}
For \: \: \: the \: \: \: Data: \: \: \: \hspace{1.5cm}<N> \: = \: 2.826  (P_{Lab})^{0.226} 
\end{equation}
	
	\end{center}


	\begin{center}
	\begin{equation}
	For \: \: \: the  \: \: \: Tsallis \: \: Model:\: \: \: \: \: <N>  \: =  \: 2.708 (P_{Lab})^{0.274}
	\end{equation}
	
	\end{center}


For $p$-Em;
\begin{center}
\begin{equation}
For \: \: \: the \: \: \: Data: \: \: \: \hspace{1.5cm}<N> \: = \: 2.281  (P_{Lab})^{0.331} 
\end{equation}
	
	\end{center}


	\begin{center}
	\begin{equation}
	For \: \: \: the  \: \: \: Tsallis \: \: Model:\: \: \: \: \: <N>  \: =  \: 2.276 (P_{Lab})^{0.335}
	\end{equation}
	
	\end{center}

In the case of $\pi^-$-Em interactions,  the average multiplicity predicted by the Tsallis model at $P_{Lab}$ = 525 GeV  is found to be, $<N>_{TS}$ = 15.36 $\pm$ 0.80  compared to the  experimental values of $<N>_{\pi- Em}$ = 15.93 $\pm$ 0.22 whereas in case of proton emulsion interactions the $<N>$ using the Tsallis model at $P_{Lab}$ = 800 GeV is found to be $<N>_{TS}$ = 19.59 $\pm$ 0.63 compared to the experimental value of $<N>_{p- Em}$ = 20.02 $\pm$ 0.29.

\subsection{Moments}

	The Tsallis model has been used to calculate the moments in order to study the correlation between the particles produced in the final state of hadron-nucleus interactions. The probability distribution function is calculated from the Tsallis model using equations~(6) and (7) and then fitting is performed  on experimental  multiplicity data  at each of the given energies.~The multiplicity distribution obtained from the Tsallis model is then used to calculate the moments of the distribution using equations ~(2)~and~(3) The dependence of  $C_{\alpha}$ and $F_{\alpha}$ moments on the  energy $P_{Lab}$   for $\pi^-$-Em and  $p$- Em data are shown in figures 3 and 4 respectively. The normalised and factorial moments from the data are compared with the moments obtained from the Tsallis model and are listed in Table II and III. Both the moments $C_{\alpha}$ and $F_{\alpha}$ obtained from the Tsallis model are found to be in good agreement with the experimental data. It is observed that the $F_{\alpha}$ moments in each case is greater than unity, confirming the correlation between the produced particles. It has been found that the moments $C_{\alpha}$ and $F_{\alpha}$ remains constant with increase in energy for  $\pi^-$-Em and  $p$- Em interactions. The independence of moments on $P_{Lab}$ clearly indicates that the KNO scaling holds good at the lower energies.

\section{Conclusions}
Detailed analysis of the experimental data on multiplicity distributions of particles produced in the interactions of proton with emulsion nuclei at incident energies between $P_{Lab}$ = 27 GeV to 800 GeV and interaction of pion with emulsion nuclei at $P_{Lab}$ = 50, 200, 340 and 525 GeV has been done.  The study of multiplicity distributions, their moments and dependence of average multiplicity on the energy helps us in understanding the process and readable features of particle production mechanism at higher energies. Experimentally the particles produced in the high energy interactions are found to be correlated.  The random and  irregular cascading processes that occur during the particle production cause the dynamical fluctuations. These dynamical fluctuations are responsible for the correlations between the produced particles. The correlation between the particles can be  perceived from the factorial moments. If the particle production is independent then factorial moments are always unity but on other hand if particles are correlated then the factorial moments are greater than unity. The normalized moments are being used  to study the holding or breaking of KNO scaling. The  energy independence of these moments signifies the upholding of  KNO scaling  whereas energy dependence of these moments reflects the KNO scaling violation. In the current study the average multiplicity, higher order moments are determined by using the Tsallis model. The values obtained from the Tsallis model are found to be in good agreement with the experimental data.  The study of Tsallis model at these energies clealry supports the upholding of KNO scaling and the correlations between the particles as observed in the experimental values. The success of the Tsallis model at both lower and higher energies makes it better and preferable choice  for studing the particle production mechanism.  In future It will be intriguing to analyse the Tsallis model on the heavy-ion data at higher enegies (at $\sqrt{s}$ $>$ 13 TeV, LHC) to explore the new aspects of particle production mechanism.

\begin{figure}[ht]
\includegraphics[width=5.0 in, height =3.0 in]{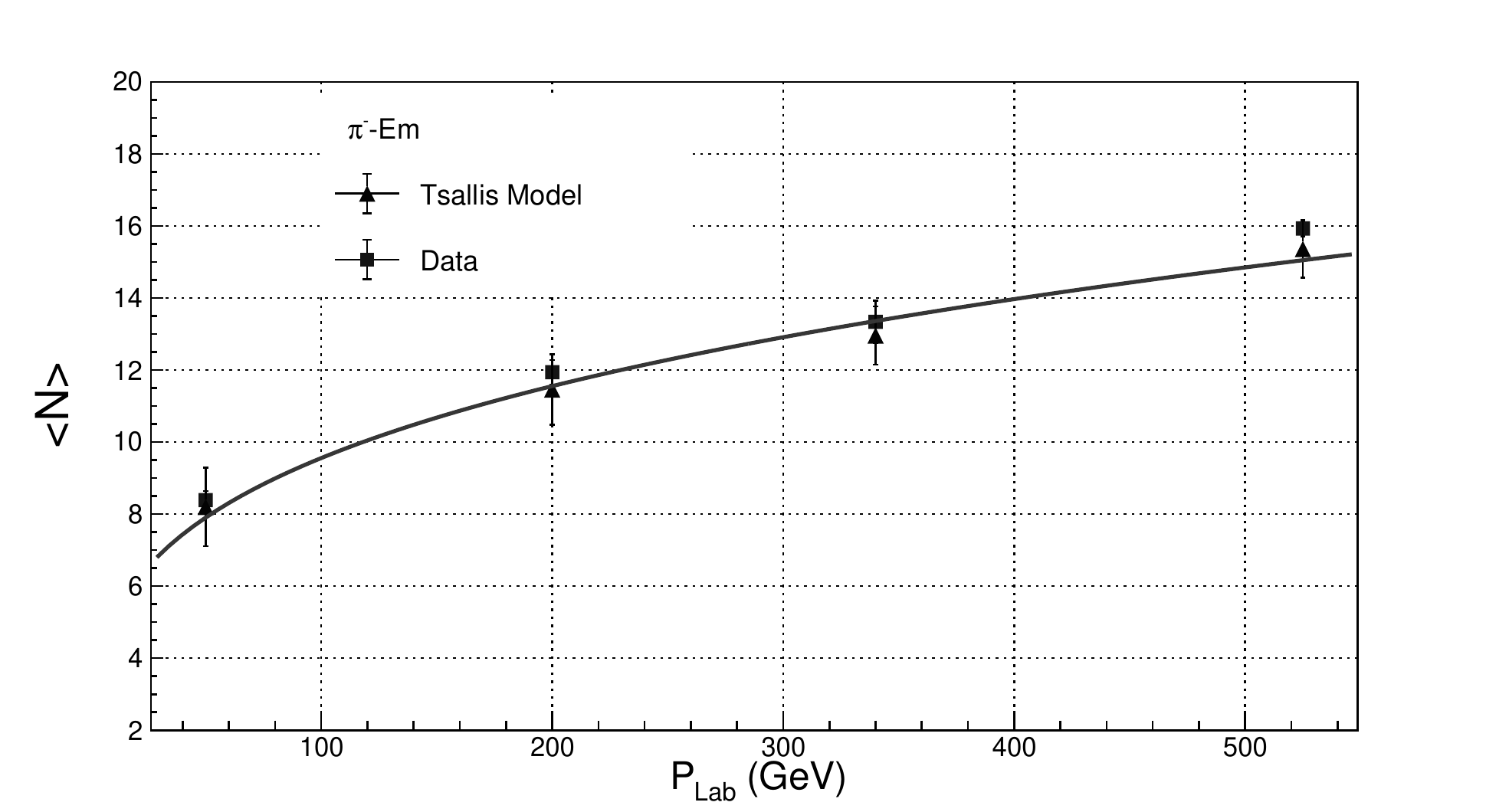}
\caption{Dependence of the average multiplicity $<N> $ on  energy, $P_{Lab}$ for $\pi^-$-Em interactions and comparison with the experimental values. The solid line corresponds to the equation (19) }
\end{figure}

\begin{figure}[ht]
\includegraphics[width=5 in, height =3.0 in]{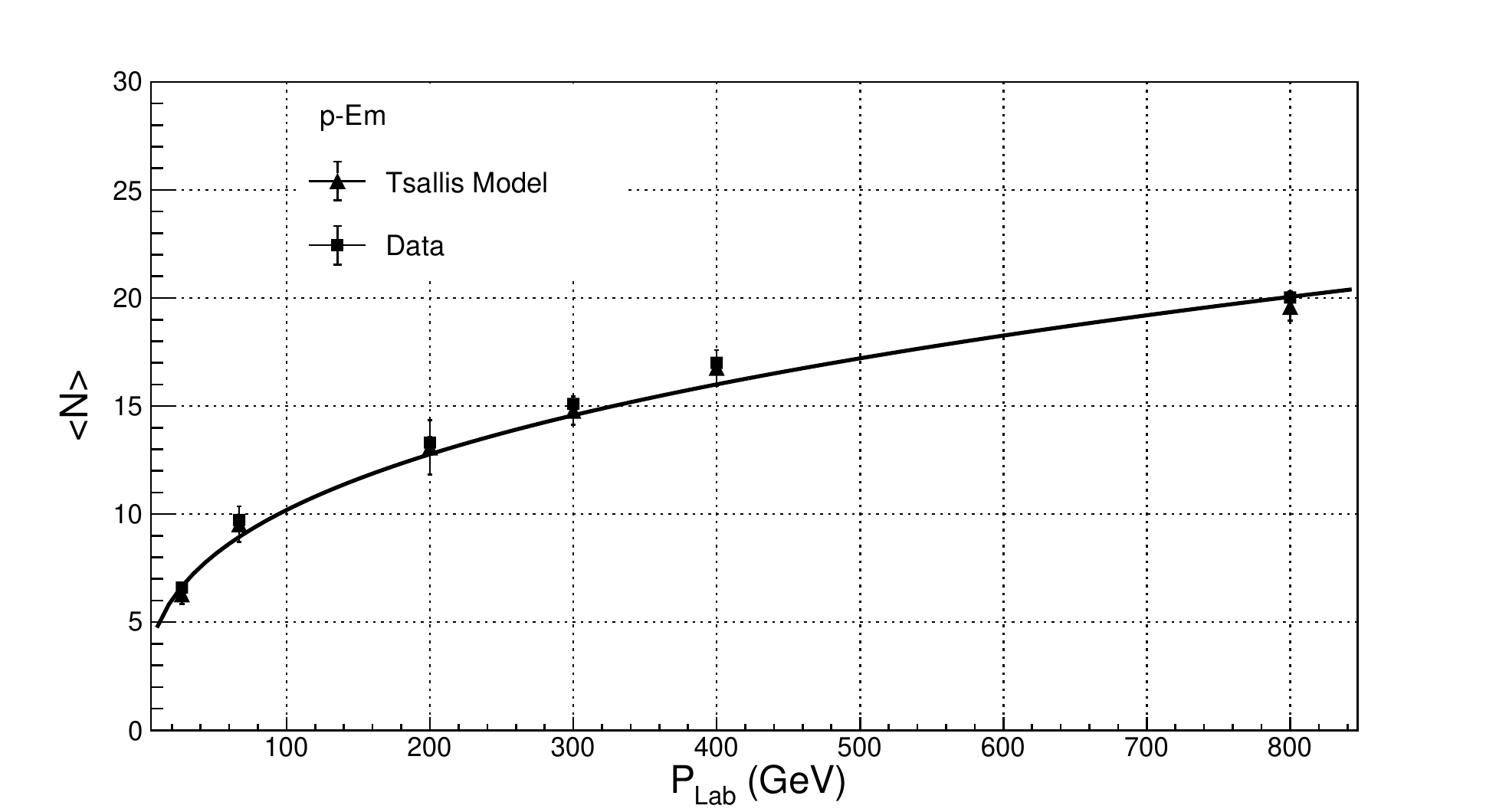}
\caption{Dependence of the average multiplicity $<N> $ on  energy, $P_{Lab}$ for  $p$-Em interactions and comparison with the experimental values. The solid line corresponds to the equation (21) }
\end{figure}

\begin{figure}[ht]
\includegraphics[width=5 in, height =3.5 in]{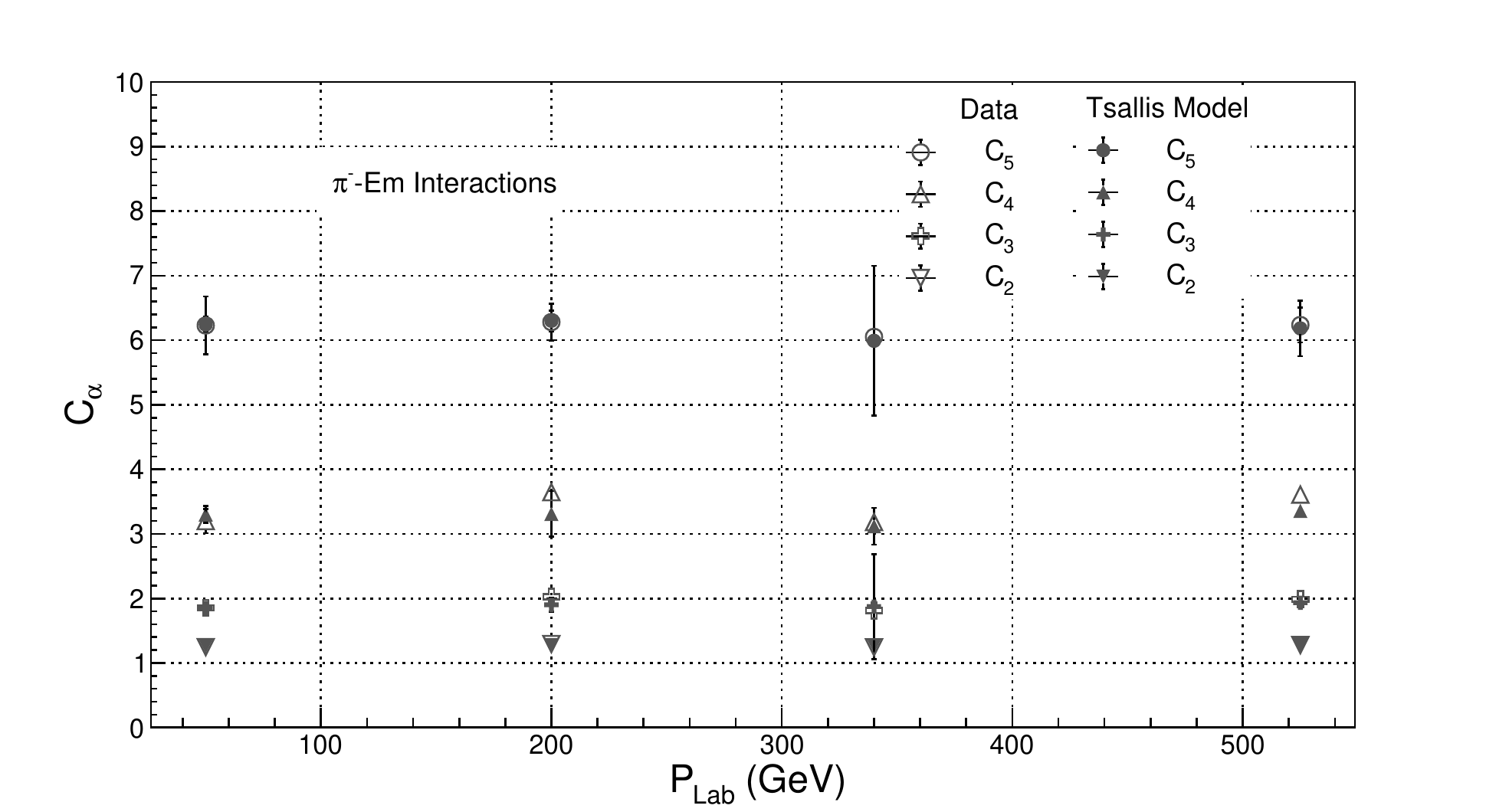}
\includegraphics[width=5 in, height =3.5 in]{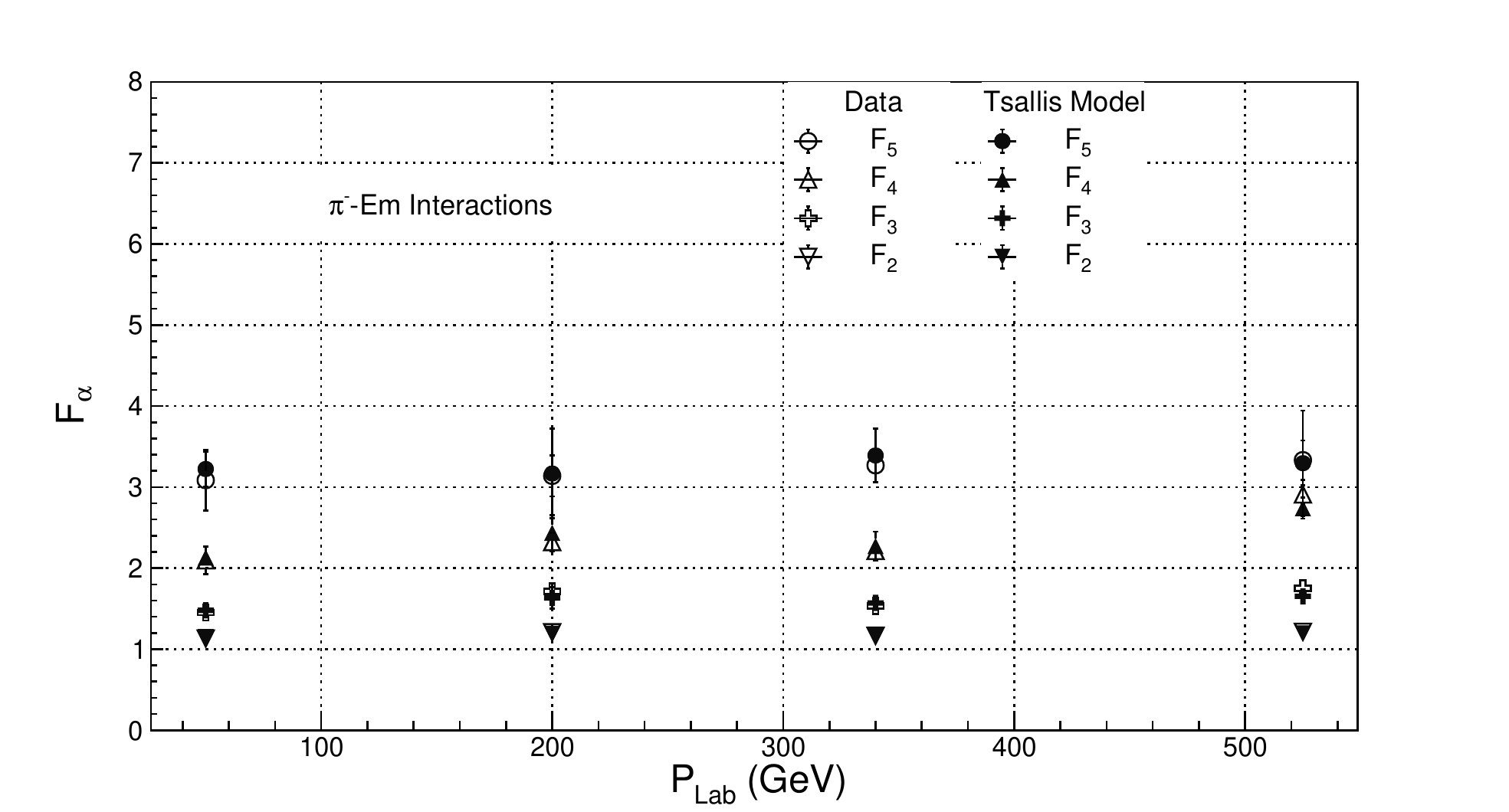}
\caption{Dependence of $C_{\alpha}$ and $F_{\alpha}$ moments calculated from the Tsallis model on  energy $P_{Lab}$ for $\pi ^-$-Em interactions and comparison with the experimental data.}
\end{figure}

\begin{figure}[ht]
\includegraphics[width=5.0 in, height =3.5 in]{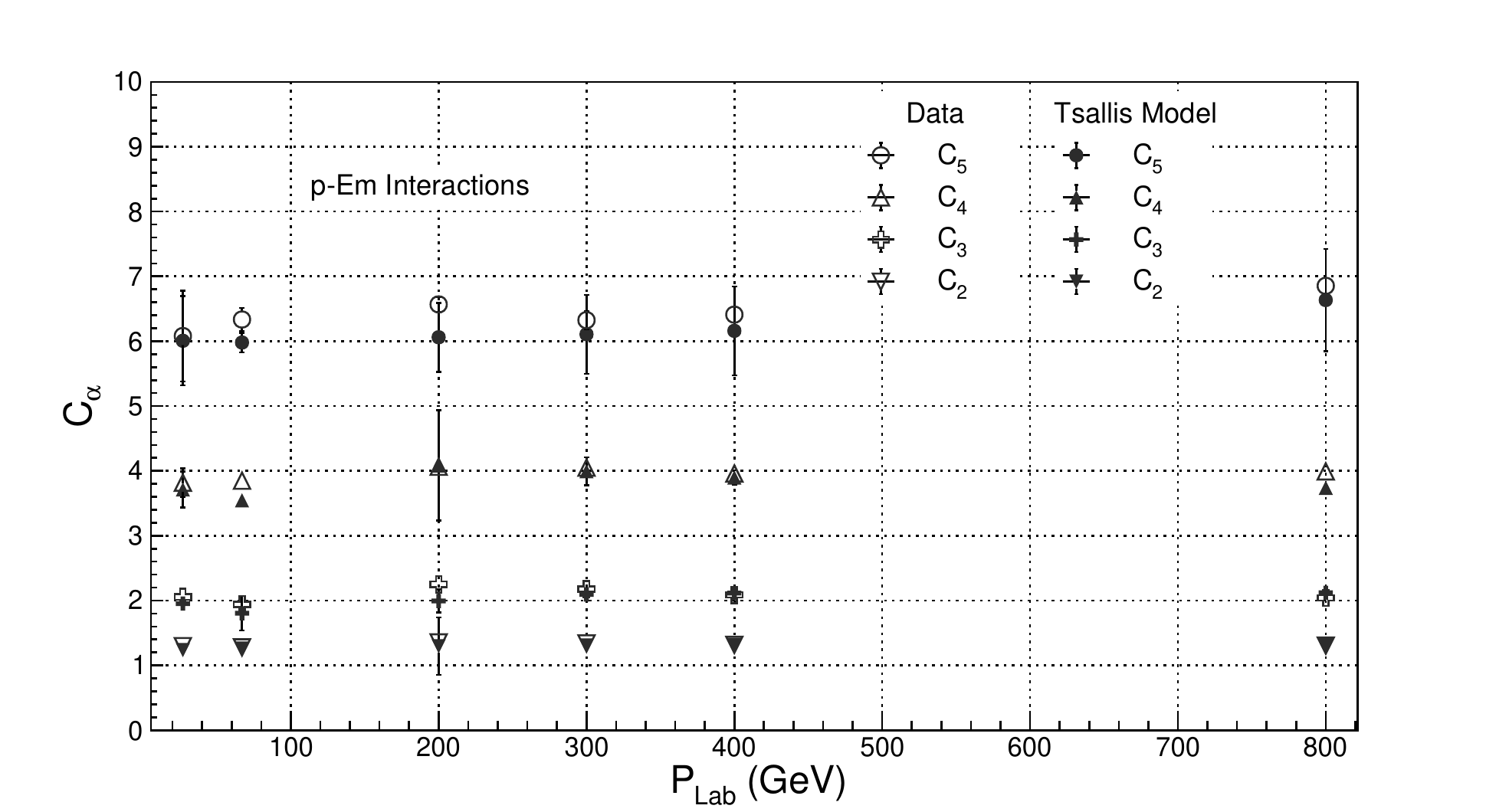}
\includegraphics[width=5.0 in, height =3.5 in]{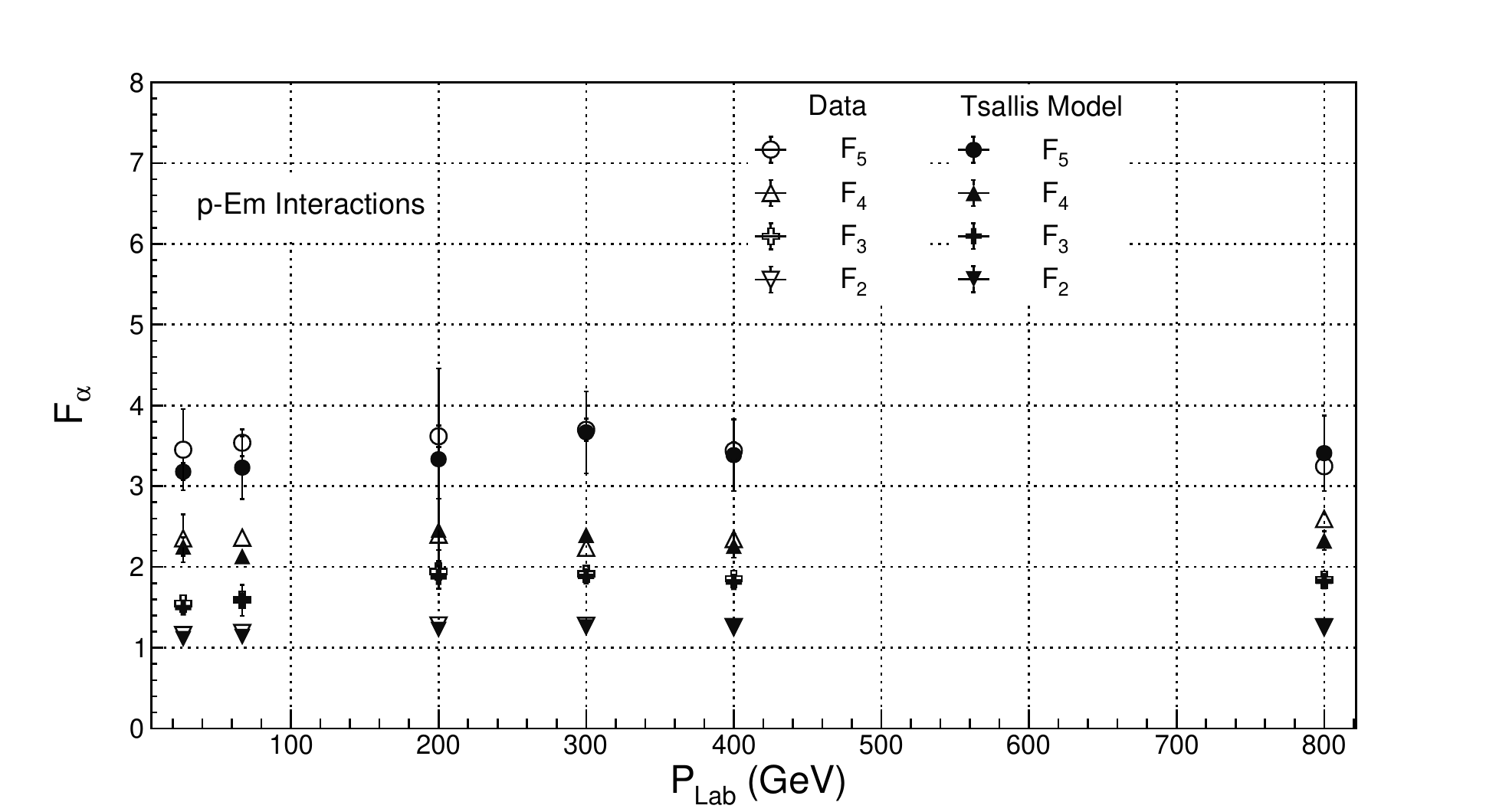}
\caption{Dependence of $C_{\alpha}$ and $F_{\alpha}$ moments calculated from the Tsallis model on  energy $P_{Lab}$ for $p$-Em interactions and comparison with the experimental data.}
\end{figure}

\small
\begin{table}[pt]
\tbl{Average charged multiplicity $<N>$  values from experiment as well as from the Tsallis model at various energies in $\pi^-$-Em and $p$-Em interactions.}
{\begin{tabular}{|c|c|c|c|c|c|c|c|}
\hline



 Energy &  Experiment & \multicolumn{2}{|c|}{Tsallis Model}\\\hline
 &    &   & \\
 $P_{Lab}$ (GeV) &  Average Multiplicity  & Average  Multiplicity   & Tsallis $q$ parameter \\
  &   $<N>$ &   $<N>$ & $q$  \\
  & & &\\\hline
  \multicolumn{4}{|c|}{}  \\ 
  			 \multicolumn{4}{|c|}{$\pi^{-}$-Em}  \\       
			  \multicolumn{4}{|c|}{}    \\\hline 

     50   &	8.39 $\pm$ 0.25   & 8.20 $\pm$ 1.09 	&1.006 $\pm$ 0.004  \\\hline
 
		200&11.94 $\pm$ 0.34 & 11.46 $\pm$ 0.98 &1.149 $\pm$ 0.017 \\\hline
		340&13.34 $\pm$ 0.59& 12.96 $\pm$ 0.81&1.069 $\pm$ 0.009 \\\hline
		525&15.93 $\pm$ 0.22& 15.36 $\pm$ 0.80&1.258 $\pm$ 0.020 \\\hline
		
		  \multicolumn{4}{|c|}{}  \\ 
  			 \multicolumn{4}{|c|}{$p$-Em}  \\       
			  \multicolumn{4}{|c|}{}    \\\hline 
		27&6.60 $\pm$ 0.10& 6.29 $\pm$ 0.44& 1.001 $\pm$ 0.002 \\\hline
		67&9.73 $\pm$ 0.23& 9.53 $\pm$ 0.83&1.034 $\pm$ 0.014\\\hline
		200&13.31 $\pm$ 0.28 & 13.09 $\pm$ 1.26& 1.149 $\pm$ 0.241 \\\hline
		300&15.10 $\pm$ 0.20& 14.74 $\pm$ 0.65& 1.369 $\pm$ 0.031\\\hline
		400&17.00 $\pm$ 0.21& 16.77 $\pm$ 0.82&1.464 $\pm$ 0.018\\\hline
		800&20.02 $\pm$ 0.29& 19.59 $\pm$ 0.63&1.633 $\pm$ 0.026\\\hline

\end{tabular}}
\end{table}

\begin{table}[htbp]
\tbl{Comparison of  reduced moments $C_{\alpha}$  from the experimental data and the Tsallis model at various energies in $\pi^-$-Em and $p$-Em interactions. }
{\begin{tabular}{|c|cccc|cccc|}
\hline
 Energy &  \multicolumn{8}{|c|}{Reduced Moments} \\\hline
    &\multicolumn{4}{|c|}{Experimental Values } &  \multicolumn{4}{|c|}{Tsallis Model }  \\\cline{2-9}
 $P_{Lab}$  &	&  &			 					 &	  &      & &&\\     
   	(GeV) &  $C_2$  & $C_3$ & $C_4$  &  $C_5$ & $C_2$ & $C_3$ & $C_4$& $C_5$\\
  &	&  			 &					 &	  &      & &&\\\hline
  \multicolumn{9}{|c|}{}  \\ 
  			 \multicolumn{9}{|c|}{$\pi^{-}$-Em}  \\       
			  \multicolumn{9}{|c|}{}    \\\hline

    50	&1.246 $\pm$ 0.020	&1.854 $\pm$ 0.063	&3.198 $\pm$ 0.188	&6.228 $\pm$ 0.449	& 1.262 $\pm$ 0.011	& 1.852 $\pm$ 0.040	& 3.301 $\pm$ 0.131 & 6.242 $\pm$ 0.118	\\\hline

200	&1.294 $\pm$ 0.012	&2.024 $\pm$ 0.044	&3.646 $\pm$ 0.117	&6.281 $\pm$ 0.282 &1.275 $\pm$ 0.024&1.901 $\pm$ 0.107&3.311 $\pm$ 0.351&6.301 $\pm$ 0.160	\\\hline

340	&1.243 $\pm$ 0.007	&1.815 $\pm$ 0.020	&3.183 $\pm$ 0.034	&6.052 $\pm$ 0.094	&1.263 $\pm$ 0.021&1.874 $\pm$ 0.814&3.119 $\pm$ 0.285&5.991 $\pm$ 1.160 \\\hline

525	&1.278 $\pm$ 0.013	&1.986 $\pm$ 0.046	&3.609 $\pm$ 0.115	&6.236 $\pm$ 0.269	&1.284 $\pm$ 0.019&1.934 $\pm$ 0.013&3.357 $\pm$ 0.027&6.183 $\pm$ 0.431\\\hline 
   \multicolumn{9}{|c|}{}  \\ 
  			 \multicolumn{9}{|c|}{$p$-Em}  \\       
			  \multicolumn{9}{|c|}{}    \\\hline   
			  
  27	&1.297 $\pm$ 0.015	&2.055 $\pm$ 0.065	&3.817 $\pm$ 0.221	&6.080 $\pm$ 0.700	&1.237 $\pm$ 0.052&1.953 $\pm$ 0.095& 3.714 $\pm$ 0.276& 6.007 $\pm$ 0.689\\\hline
  
67	&1.278 $\pm$ 0.009	&1.943 $\pm$ 0.030	&3.849 $\pm$ 0.077	&6.335 $\pm$ 0.176       &1.258 $\pm$ 0.026&1.804 $\pm$ 0.265& 3.545 $\pm$ 0.036& 5.980 $\pm$ 0.152\\\hline	

200	&1.354 $\pm$ 0.009	&2.251 $\pm$ 0.028	&4.064 $\pm$ 0.065	&6.568 $\pm$ 0.119	&1.299 $\pm$ 0.443&1.993 $\pm$ 0.178& 4.083 $\pm$ 0.850& 6.061 $\pm$ 0.532 \\\hline    

300	&1.338 $\pm$ 0.008	&2.175 $\pm$ 0.026	&4.050 $\pm$ 0.062	&6.325 $\pm$ 0.140	&1.307 $\pm$ 0.014&2.094 $\pm$ 0.073&  3.995 $\pm$ 0.214& 6.107 $\pm$ 0.609 \\\hline 

400	&1.317 $\pm$ 0.007	&2.087 $\pm$ 0.021	&3.959 $\pm$ 0.051	&6.410 $\pm$ 0.113      &1.319 $\pm$ 0.029&2.114 $\pm$ 0.055& 3.899 $\pm$ 0.110& 6.160 $\pm$ 0.685\\\hline	

800	&1.306 $\pm$ 0.007	&2.041 $\pm$ 0.017	&3.992 $\pm$ 0.031	&6.854 $\pm$ 0.038	&1.312 $\pm$ 0.092&2.120 $\pm$ 0.063& 3.739 $\pm$ 0.087&6.635 $\pm$ 0.790  \\\hline

	\end{tabular}}
\vspace{0.3 cm}
\end{table}

\begin{table}[htbp]
\tbl{Comparison of factorial moments $F_{\alpha}$  from the experimental data and the Tsallis model at various energies in $\pi^-$-Em and $p$-Em interactions. }
{\begin{tabular}{|c|cccc|cccc|}
\hline
 Energy &  \multicolumn{8}{|c|}{Factorial Moments} \\\hline
    &\multicolumn{4}{|c|}{Experimental Values } &  \multicolumn{4}{|c|}{Tsallis Model }  \\\cline{2-9}
 $P_{Lab}$  &	&  &			 					 &	  &      & &&\\     
   	(GeV) &  $F_2$  & $F_3$ & $F_4$  &  $F_5$ & $F_2$ & $F_3$ & $F_4$& $F_5$\\
  &	&  			 &					 &	  &      & &&\\\hline
  \multicolumn{9}{|c|}{}  \\ 
  			 \multicolumn{9}{|c|}{$\pi^{-}$-Em}  \\       
			  \multicolumn{9}{|c|}{}    \\\hline

50	&1.131 $\pm$ 0.019	&1.453 $\pm$ 0.066	&2.095 $\pm$ 0.169	&3.085 $\pm$ 0.372 & 1.148 $\pm$ 0.016& 1.479 $\pm$ 0.036& 2.130 $\pm$ 0.080 & 3.223 $\pm$ 0.218\\\hline

200	&1.209 $\pm$ 0.015	&1.711 $\pm$ 0.050	&2.321 $\pm$ 0.118	&3.138 $\pm$ 0.254  &1.189 $\pm$ 0.017       &1.633 $\pm$ 0.131&2.437 $\pm$ 0.217&3.167 $\pm$ 0.553\\\hline

340	&1.165 $\pm$ 0.008	&1.536 $\pm$ 0.021	&2.212 $\pm$ 0.044	&3.268 $\pm$ 0.077 &1.179 $\pm$ 0.016       &1.571 $\pm$ 0.016&2.272 $\pm$ 0.179&3.390 $\pm$ 0.332\\\hline

525	&1.215 $\pm$ 0.013	&1.751 $\pm$ 0.046	&2.909 $\pm$ 0.113	&3.333 $\pm$ 0.242  &1.194 $\pm$ 0.007       &1.652 $\pm$ 0.015&2.741 $\pm$ 0.131&3.295 $\pm$ 0.649\\\hline
 
   \multicolumn{9}{|c|}{}  \\ 
  			 \multicolumn{9}{|c|}{$p$-Em}  \\       
			  \multicolumn{9}{|c|}{}    \\\hline   
			  
27	&1.157 $\pm$ 0.018	&1.549 $\pm$ 0.069	& 2.354 $\pm$ 0.295	&3.452 $\pm$ 0.503 & 1.108 $\pm$ 0.016      &1.498 $\pm$ 0.049&2.252 $\pm$ 0.113&3.179 $\pm$ 0.105\\\hline

67	&1.183 $\pm$ 0.009	&1.596 $\pm$ 0.036	& 2.363 $\pm$ 0.079	&3.538 $\pm$ 0.164 &1.133 $\pm$ 0.009 &1.585 $\pm$ 0.192&2.133 $\pm$ 0.091&3.230 $\pm$ 0.387\\\hline

200	&1.276 $\pm$ 0.011	&1.946 $\pm$ 0.033	& 2.401 $\pm$ 0.072	&3.618 $\pm$ 0.135 &1.221 $\pm$ 0.034&1.878 $\pm$ 0.148&2.460 $\pm$ 0.385& 3.334 $\pm$ 1.125\\\hline

300	&1.271 $\pm$ 0.010	&1.915 $\pm$ 0.028	& 2.238 $\pm$ 0.067	&3.698 $\pm$ 0.139 &1.252 $\pm$ 0.081&1.885 $\pm$ 0.055&2.398 $\pm$ 0.024& 3.667 $\pm$ 0.507\\\hline

400	&1.256 $\pm$ 0.008	&1.853 $\pm$ 0.023	& 2.346 $\pm$ 0.053	&3.440 $\pm$ 0.101 &1.259 $\pm$ 0.009&1.814 $\pm$ 0.032&2.262 $\pm$ 0.145& 3.385 $\pm$ 0.444\\\hline

800	&1.254 $\pm$ 0.008	&1.843 $\pm$ 0.019	& 2.594 $\pm$ 0.036	&3.247 $\pm$ 0.050 &1.267 $\pm$ 0.010&1.823 $\pm$ 0.049&2.327 $\pm$ 0.117& 3.409 $\pm$ 0.467\\\hline

	\end{tabular}}
\vspace{0.3 cm}
\end{table}


\newpage
\begin{thebibliography}{}

\bibitem{int1}  V. V. Anisovich, M. N. Kobrinsky, Yu. M. Shabelski and J. Nyiri,  Quark Model and High Energy Collisions, Second Edition, \textit{World Scientific Pub Co Inc}, ( 2004)


\bibitem{hh2} R. P. Feynman, \textit{ Phys. Rev. Lett. }\textbf{23},  (1969) 1415

\bibitem{ee3} I. Bediaga, E.M.F. Curado and J.M. de Miranda, \textit{ Physica A} \textbf{286},  (2000) 156

\bibitem{ha4} N. Dobrotin, \textit{ICRC} \textbf{10},  (1977) 456

\bibitem{fragmentation5} B. Webber, \textit{ Int. J. Mod. Phys. A} \textbf{15},  (2000)  577

\bibitem{e0}  G. Wilk and  Z. Włodarczyk,	\textit{Int. J. Mod. Phys. A} \textbf{33},  (2018) 1830008

\bibitem{qgp6}  H. Bohr and H. B. Nielsena,  \textit{ Nucl. Phys. B.} \textbf{128 (2)}, (1977)  275

\bibitem{md7}  C. Fuglesang, \textit{Multiparticle Dynamic}, edited by A.
Giovannini and W. Kittel, World Scientific, Singapore,  (1997) 193

 
  
\bibitem{poisson8} D. J. Mangeol, arXiv:hep-ex/0110029

\bibitem{cummulants9}  A. Giovannini, S. Lupia, and R. Ugoccioni, \textit{ Phys. Lett. B} \textbf{374}, (1996) 231


\bibitem{models10} Z. Koba, D. Weingarten,  \textit{Lett. al Nuov. Cim.} \textbf{8}, (1973)  303 

\bibitem{pt11} G. Wilk and Z. Wlodarczyk, \textit{Physica A} \textbf{305} (2002) 227.


\bibitem{Tsallis12} C. Tsallis,  \textit{J. Stat. Phys.} \textbf{52}, (1988) 479 

\bibitem{kodoma13} C.E. Aguiar and T. Kodama, \textit{Physica A} \textbf{320},  (2003) 371

\bibitem{s1} S. Sharma, M. Kaur and S. Kaur, \textit{Int. J. Mod. Phys. E} \textbf{25} (2016) 1650041
\bibitem{s2} S. Sharma, M. Kaur and S. Thakur, \textit{Phys. Rev. D} \textbf{95} (2017) 114002 
\bibitem{s3} S. Sharma  and  M. Kaur, \textit{Phys. Rev. D} \textbf{98} (2018) 034008 
\bibitem{s4} S. Sharma, M. Kaur and S. Thakur, \textit{Int. J. Mod. Phys. E} \textbf{27} (2018) 1850101

\bibitem{mypaper17} S. Sharma, M. Kaur and S. Thakur,  \textit{Int. J. Mod. Phys. E} \textbf{26}, (2017) 1750006 

\bibitem{nbd14} J. F. Lawless,  \textit{CJS} \textbf{15},  (1987) 209

\bibitem{gamma15} K. Urmossy, G. G. Barnafoldi and T. S. Biro, \textit{Phys. Lett. B} \textbf{701},  (2011) 111

\bibitem{shifted gamma16} K. Urmossy, G.G. Barnafoldi and T. S. Biro, \textit{Phys. Lett. B} \textbf{718},  (2012) 125


\bibitem{generating18}  A. Capella, I. M. Dremin, V. A. Nechitailo and J. T. T. Van, \textit{Z.Phys. C} \textbf{75}, 89 (1997)

\bibitem{kno19} Z. Koba, H. B. Nielsen, P. Olesen, \textit{ Nucl. Phys. B }\textbf{40},  (1972)  317

\bibitem{m20} A. H. Mueller. \textit{Phys. Rev. D} \textbf{4},  (1971) 150


\bibitem{m21}  P. Zhuang, \textit{Phys. Rev. C} \textbf{66}, (2002) 064901

\bibitem{m} I. M. Dremin and J. W. Gary,  \textit{Phys. Rept. }\textbf{349} (2001) 301 
\bibitem{m22}  S. Radka, T. Boris, and B. Marcus, \textit{Phys. Rev. C }\textbf{98},  (2018) 064907


\bibitem{gibbs23} R. Hagedorn, \textit{Riv. del Nuovo Cimento} \textbf{6},  (1983) 1-50


\bibitem{error24} J. F. Grosse-Oetringhaus, K. Reygers, \textit{J. Phys. G} \textbf{37},  (2010) 083001


\bibitem{fixexperiment25} G. P. S. Occhialini, C. F. Powell, \textit{ Nature} \textbf{159},  (1947) 453 


\bibitem{meann26}  E. Fermi et al.,  \textit{Prog. Theor. Phys.} \textbf{5},  (1950) 570

\bibitem{meann27} H. Satz, \textit{Current Induced Reactions} \textbf{56}, (1975) 49 

\end{thebibliography}
\end{document}